\documentclass[aps,apl,twocolumn,superscriptaddress]{revtex4-1}
\usepackage{amsfonts}
\usepackage{amsmath}
\usepackage{amssymb}
\usepackage{graphicx}

\begin{document}


\title{Sub-nanosecond switching in a cryogenic spin-torque spin-valve memory element with a dilute permalloy free layer}

\author{L. Rehm}
\email{laura.rehm@nyu.edu}
\affiliation{Center for Quantum Phenomena, Department of Physics, New York University, New York, NY 10003, USA}
\author{V. Sluka}
\email{volker.sluka@julumni.fz-juelich.de} 
\affiliation{Center for Quantum Phenomena, Department of Physics, New York University, New York, NY 10003, USA}
\author{G. E. Rowlands}
\affiliation{Raytheon BBN Technologies, Cambridge, MA 02138, USA}
\author{M.-H. Nguyen}
\affiliation{Raytheon BBN Technologies, Cambridge, MA 02138, USA}
\author{T. A. Ohki}
\affiliation{Raytheon BBN Technologies, Cambridge, MA 02138, USA}
\author{A. D. Kent}
\email{andy.kent@nyu.edu}
\affiliation{Center for Quantum Phenomena, Department of Physics, New York University, New York, NY 10003, USA}

\date{\today}

\begin{abstract}
We present a study of the pulsed current switching characteristics of spin-valve nanopillars with in-plane magnetized dilute permalloy and undiluted permalloy free layers in the ballistic regime at low temperature. The dilute permalloy free layer device switches much faster: the characteristic switching time for a permalloy free (Ni\textsubscript{0.83}Fe\textsubscript{0.17}) layer device is 1.18\,ns, while that for a dilute permalloy ([Ni\textsubscript{0.83}Fe\textsubscript{0.17}]\textsubscript{0.6}Cu\textsubscript{0.4}) free layer device is 0.475\,ns. A ballistic macrospin model can capture the data trends with a reduced spin-torque asymmetry parameter, reduced spin polarization and increased Gilbert damping for the dilute permalloy free layer relative to the permalloy devices. Our study demonstrates that reducing the magnetization of the free layer increases the switching speed while greatly reducing the switching energy and shows a promising route toward even lower power magnetic memory devices compatible with superconducting electronics.
\end{abstract}


\pacs{}

\maketitle

There is a growing interest in spin-transfer devices that work in a cryogenic environment, such as for use in superconducting logic and circuits~\cite{Holmes2013}. While past low temperature memory efforts combined, for example, Josephson and complementary metal-oxide semiconductor devices in hybrid circuits or  explored circuits that stored magnetic flux quanta in superconducting loops~\cite{van201364,nagasawa1995380}, these approaches did not simultaneously offer high speed, low power, and scalability. Spin-transfer torque (STT) driven magnetic memory elements are known to be non-volatile, fast, and energy efficient~\cite{Slonczewski1996,Berger1996}, but so far, they are almost exclusively being developed and tested for commercial applications~\cite{KentWorledge2015}, which require operation at and above room temperature. Cryogenic operation with superconducting circuits change device and material requirements. For example, the magnetic anisotropy energy barrier that stabilizes the magnetic states and permits long-term data retention can be greatly reduced. Large magnetoresistance also may not be essential given the sensitivity of superconducting circuits and the reduced thermal noise at low temperature. This makes it promising to study all metallic spin-valve structures, both due to their low impedance and potential for fast switching~\cite{Bedau2010,Rowlands2019}.

Spin-transfer induced magnetization switching is fundamentally based on the transfer of angular momentum between itinerant electrons and background magnetization. Switching thus requires that the number of electrons that flow through a circuit to be of order of the number of elemental magnetic moments (or spins) in the free layer~\cite{Sun2000}. This requirement sets the order of magnitude of the product of the current and the switching time (which is proportional to the total number of charges transmitted) in what is known as the ballistic limit, the short-pulse-time limit (typically pulse durations less than several nanoseconds) in which thermal energy has a minimal effect on the switching dynamics~\cite{Liu2014}. Reducing the magnetization density is thus expected to reduce the switching current. It is also expected to increase the switching speed and thus reduce the switching energy, which is a product of the power supplied and the time the device is energized.

In this article we test this hypothesis by comparing the switching characteristics of spin-valve nanopillars with in-plane magnetized dilute permalloy and undiluted permalloy free layers, but otherwise the same layers, nanopillar shape and size. In both cases the layer stacks are deposited on a Niobium (Nb) bottom electrode to show that integration with superconducting materials is practical. We characterize the pulsed current switching thresholds in the ballistic regime for both composition free layers and find a significant decrease in the characteristic time scale from 1.18\,ns for permalloy to 0.475\,ns for the dilute permalloy free layer. A macrospin model was used to fit the switching time data with a reduced spin-torque asymmetry parameter, reduced spin polarization and increased Gilbert damping for the dilute permalloy free layer.

\begin{figure}[h]
\includegraphics[width=0.48\textwidth,keepaspectratio]{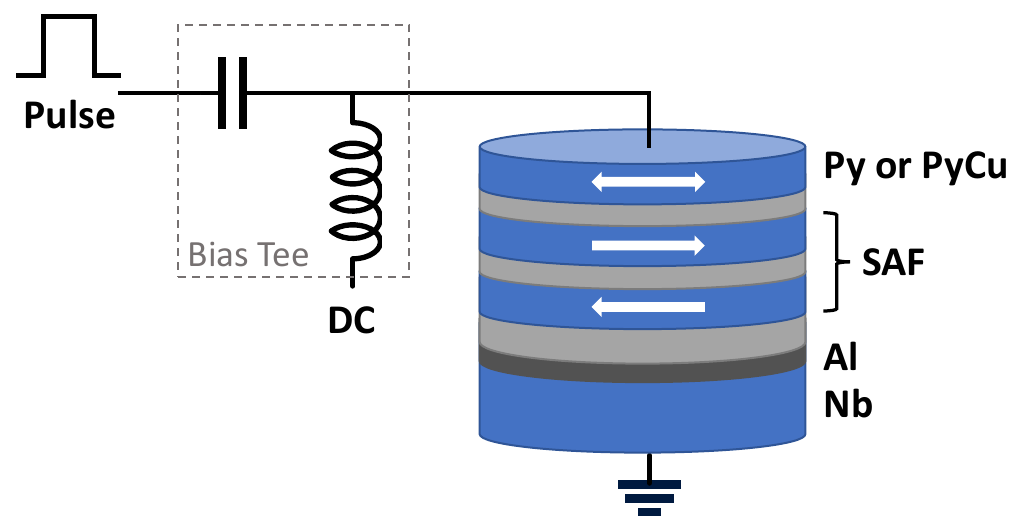}
\caption{Schematic of a spin-valve nanopillar device with an undiluted permalloy (Py) or diluted permalloy (PyCu) free layer. The write pulses \textit{I}\textsubscript{w} are applied through the capacitive port of a bias tee while the inductive port is used to read out the state of the device.}
\label{Fig:Fig1}
\end{figure}

We investigated two sets of spin-valve nanopillar devices. One with an undiluted permalloy (Ni\textsubscript{0.83}Fe\textsubscript{0.17}, denoted as Py) free layer and another with a diluted permalloy ([Ni\textsubscript{0.83}Fe\textsubscript{0.17}]\textsubscript{0.6}Cu\textsubscript{0.4}, denoted as PyCu)  free layer. The layer stacks consist of a Nb(50)/Al(8) bottom electrode layer, a CoFe(3) reference layer (RL) which is part of a synthetic antiferromagnet (SAF) CoFe(3)/Ru(0.8)/CoFe(3), and a 3\,nm thick Py or PyCu free layer (FL): Nb(50)/Al(8)/IrMn (10)/SAF/Co(0.2)/Cu(3.5)/Co(0.2)/FL, as shown in Fig.~1. The numbers in brackets are the layer thicknesses in nm. The Nb bottom electrode enables the integration with superconducting circuitry, while the Al interlayer is crucial for the properties of the magnetic stack: it wets the surface of the Nb layer and creates a smoother surface, reducing the effect of N\'eel ``orange peel'' coupling between layers~\cite{Neel1962} and effects of roughness on the magnetic switching characteristics.

\begin{figure}[h]
\includegraphics[width=0.48\textwidth,keepaspectratio]{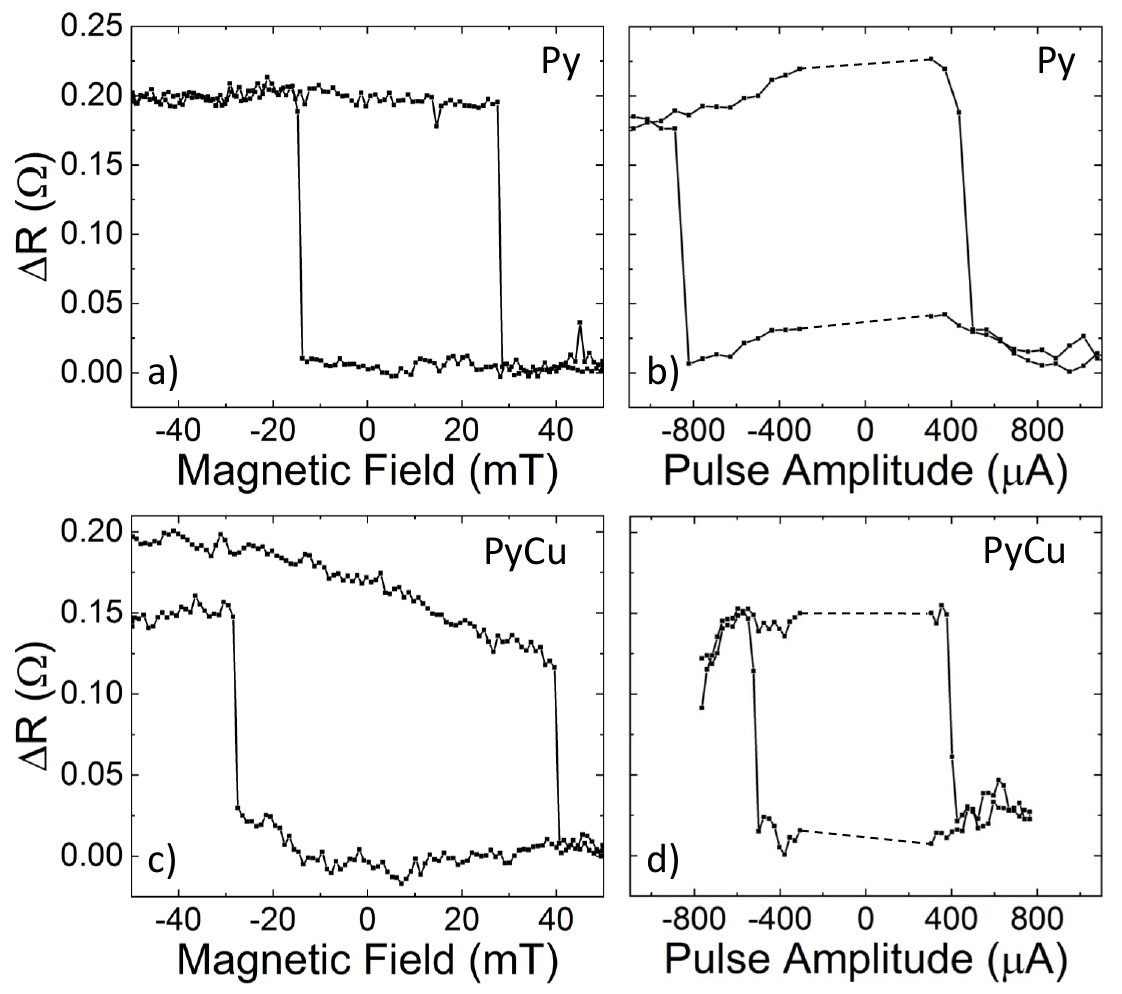}
\caption{Field- and current-induced magnetization switching of Py and PyCu free layer device at 3.2\,K.  Panels a) and c) show field-induced switching of devices with Py and PyCu free layer, respectively. The hysteresis loop shown in panel c) does not fully close due to drift in the measurement setup. The external field is applied along the easy axis of the elliptically shaped nanopillar. Panels b) and d) display current-induced switching for 10\,ns long pulses of the same set of devices. No data was taken along the dashed lines.}
\label{Fig:Fig2}
\end{figure}

Following the deposition, the wafers were annealed at 230$^\circ$C and 1\,T to set the magnetization orientation of the SAF. The annealed wafers were pattered into elliptically shaped nanopillars of various sizes using e-beam lithography and ion-milling. Here, we present results on devices with a 50\,nm $\times$ 110\,nm cross-section. The devices are characterized by measuring their field and current pulse resistance hysteresis loops at 3.2\,K. The state of the device is recorded using a lock-in technique. Small AC currents of 20 and 40\,$\mu$A are applied for the PyCu and Py free layer device, respectively. Figures~2a) and ~2c) show the minor loops of the Py and PyCu free layer device, respectively. The Py sample exhibits a resistance change between the antiparallel (AP) and parallel (P) magnetic configuration of 190\,m$\Omega$, while the PyCu free layer device exhibits a $\Delta$\textit{R} of around 120\,m$\Omega$. Both devices show a well-centered hysteresis with a small offset field of 6\,mT. Both samples also show a bistable region around zero applied current and current-induced switching with 10 ns duration current pulse with pulse amplitudes of 403 $\mu$A (for AP$\rightarrow$P switching) and -523\,$\mu$A (for P$\rightarrow$AP switching) of the PyCu free layer (Fig.~\ref{Fig:Fig2}b)) and 480 $\mu$A (AP$\rightarrow$P) and -868 $\mu$A (P$\rightarrow$AP ) for the Py free layer sample (Fig.~\ref{Fig:Fig2}d)). A difference in the P$\rightarrow$AP and AP$\rightarrow$P switching current magnitude is often observed in spin-valves and associated with spin-torque asymmetries, as discussed in Refs.~\cite{Slonczewski2002,Stiles2002,StilesM2002}.

In order to explore high speed spin-torque switching, short current pulses with durations of less than 5\,ns were used. Pulses are applied using a pulse generator (Picosecond Pulse Labs 10,070A) as well as an arbitrary waveform generator (AWG, Keysight M8190A). The first generator provides the short pulses to explore the ballistic regime, while the second generator is used to apply longer (20\,ns) duration pulses to reset the magnetization direction of the free layer. To increase the pulse amplitude resolution (below the 1\,dB resolution of the pulse generator's internal step attenuator) a voltage controlled variable attenuator (RFMD RFSA2113SB) is employed. The state of the device is determined by applying a small AC current and using a lock-in amplifier to determine the device resistance. The lock-in amplifier is operated at a 4 kHz baseband. We use a bias-tee (Picosecond Pulse Labs 5575A) to combine low-frequency measurement and high-frequency switching pulses (see Fig.~\ref{Fig:Fig1}). Two 0\,dB attenuators at the 4\,K and 50\,K stage are utilized to thermalize the center conductor of a ground signal ground (GSG) probe. A small external field (6 mT) applied along the long axis of the ellipse is used to conduct these pulse studies at the midpoint of the free layer hysteresis loop.

\begin{figure}[htp]
\includegraphics[width=0.48\textwidth,keepaspectratio]{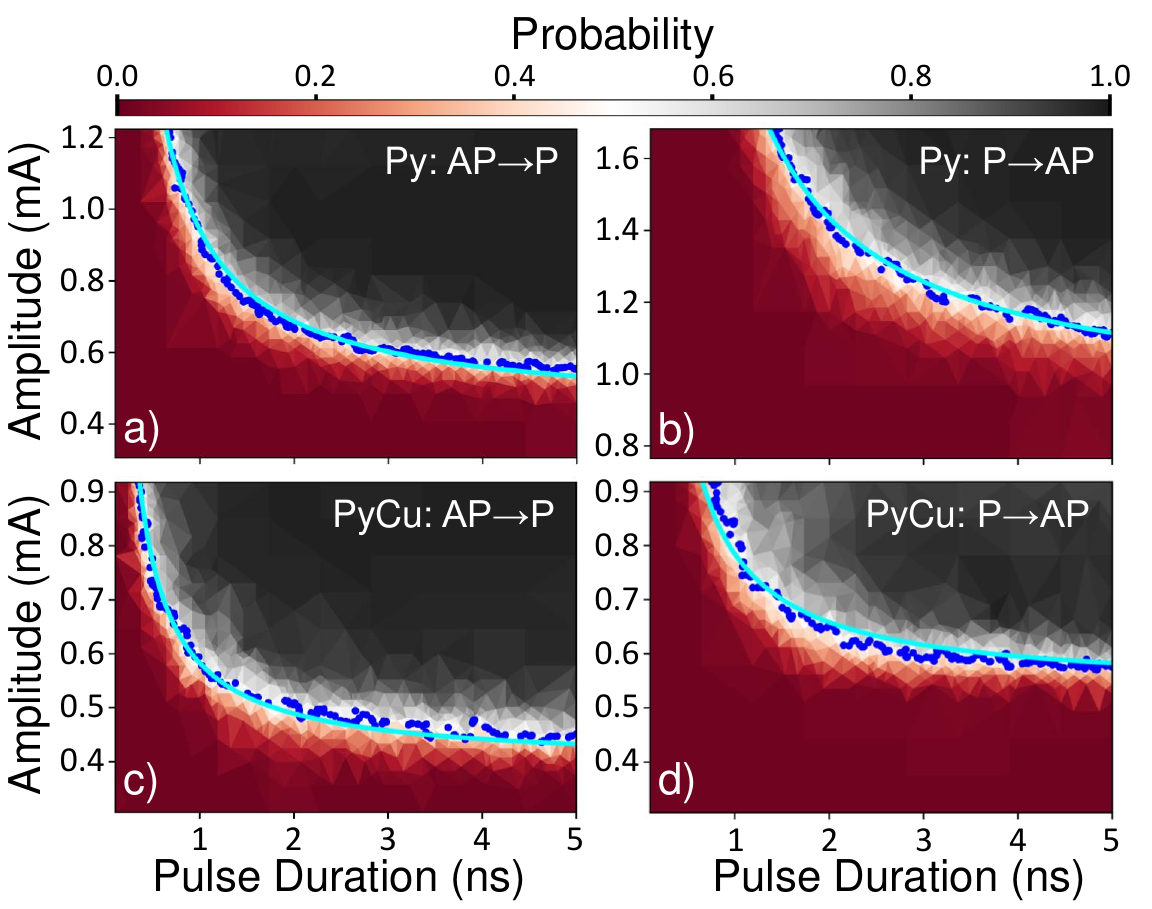}
\caption{Nanosecond pulsed current switching results at 3.2\,K.  Switching phase diagrams of a device with Py free layer, a) AP$\rightarrow$P and b) P$\rightarrow$AP, and a PyCu free layer, c) AP$\rightarrow$P and d) P$\rightarrow$AP. The color in the plot represents the switching probability, where red corresponds to 0\% and black is 100\%. The blue points represent the 50\% switching probability and the solid cyan line shows the fit to the macrospin model described in the main text.}
\label{Fig:Fig3}
\end{figure}

The measurement procedure thus consists of applying two square pulses (reset \textit{I}\textsubscript{RST} and write \textit{I}\textsubscript{w} pulses) with opposite pulse amplitudes and reads after each pulse. We start by applying a reset pulse to bring the device to a known state, either P or AP. We then verified the desired state by measuring the resistance of the device. The subsequent write pulse is applied by the pulse generator and the end state is again determined by measuring the device resistance. The whole procedure is repeated about 64 times for each write pulse amplitude and duration to determine the switching probability. We vary amplitude and duration of the write pulse to create the phase diagrams shown in Fig.~3. All the pulse measurements were performed at 3.2\,K. 

Figure~\ref{Fig:Fig3} shows the switching phase diagrams for AP$\rightarrow$P (left panels) and P$\rightarrow$AP transitions (right panels) for Py (Figs.~\ref{Fig:Fig3}(a) and (b)) and PyCu free layer device (Fig.~\ref{Fig:Fig3}(c) and (d)). The results from these samples differ significantly. For longer pulse durations, $\sim$5\,ns, switching of the PyCu free layer device occurs for lower pulse amplitudes, especially for the P$\rightarrow$AP transition (Figs.~\ref{Fig:Fig3}(d)). The PyCu free layer device also switches with high probability for shorter duration pulses than the device with the Py free layer, as seen by form of the switching boundaries (blue points in Fig.~\ref{Fig:Fig3}) for pulse durations less than 1\,ns. For the P$\rightarrow$AP direction comparatively longer pulse durations are required for switching, as discussed further below.

\begin{table*}[htp]
\caption{Fit parameters from pulsed switching measurements in the ballistic regime and corresponding spin-torque asymmetry parameter $\Lambda$ of PyCu and PyCu free layer devices. Saturation magnetization \textit{M}\textsubscript{s} for Py and PyCu layers at 3.2\,K was determined by VSM measurements.}
\centering
\label{t:Fit parameters}
\setlength{\tabcolsep}{10pt}
\begin{tabular}{lcccccc}
\noalign{\smallskip} \hline \hline \noalign{\smallskip}
Sample & \multicolumn{2}{c}{\textit{I}\textsubscript{c}($\mu$A)} & $\Lambda$ & $\tau$\textsubscript{0}(ns) & $\mu$\textsubscript{0}\textit{M}\textsubscript{s,3.2\,K}(mT)\\
 & AP$\rightarrow$P & P$\rightarrow$AP &  &  & \\
\hline
PyCu FL & 395$\pm$2 & 532$\pm$2 & 1.16 &0.475$\pm$0.007 & 240 \\
Py FL  & 432$\pm$2 & 902$\pm$3 & 1.44 &1.18$\pm$0.01 & 860 \\
\noalign{\smallskip} \hline \noalign{\smallskip}
\end{tabular}
\end{table*}

In order to understand the data trends in Fig.~\ref{Fig:Fig3} we consider a macrospin model, a simple model that provides analytic expressions for the switching times in the ballistic limit and how they vary with material and device parameters~\cite{Sun2000,Koch2004}. Since the devices are metallic spin-valves (in contrast to magnetic tunnel junctions), the spin-transfer torque angular dependence is expected to be asymmetrical, to be different for angular deviations from the P and AP states, and characterized by a parameter $\Lambda$ \cite{Slonczewski2002}, with a ratio of threshold currents given by \textit{I}\textsubscript{c}\textsuperscript{P$\rightarrow$AP}/\textit{I}\textsubscript{c}\textsuperscript{AP$\rightarrow$P} = $\Lambda$\textsuperscript{2}. Incorporating this asymmetry into a model for switching of biaxial anisotropy macrospins, and following the approach of Ref.~\cite{Koch2004}, we derive an approximate formula relating the switching speed 1/$\tau$ ($\tau$  being the switching time) to the overdrive current $I-I_c$. Due to the spin-torque asymmetry, P$\rightarrow$AP and AP$\rightarrow$P switching differ. While the relation for the former case remains the same as in Ref.~\cite{Koch2004},
\begin{equation}
\tau^{-1} = \frac{\gamma \hbar P}{4 e \mu_0 M_s V} \frac{1}{\ln\left(\frac{\pi}{2 \theta_0}\right)} \left(I - I_c^{P\rightarrow AP} \right),
\label{eq:1}
\end{equation}
for the other switching direction we have
\begin{equation}
\left(\Lambda^2 \tau\right)^{-1} = \frac{\gamma \hbar P}{4 e \mu_0 M_s V} \frac{1}{\ln\left(\frac{\pi}{2 \theta_0}\right)} \left(I - I_c^{AP\rightarrow P} \right),
\label{eq:2}
\end{equation}
where all currents are taken as positive. In these expressions $P$ is the spin polarization of the current, $M_s$ the free layer saturation magnetization, $V$ the free layer volume, $\gamma$ the gyromagnetic ratio, $\mu_0$ the vacuum permeability, $\hbar$ the reduced Planck's constant and $e$ the magnitude of the electron charge (i.e., $e>0$).  $\theta_0$ is the initial angular deviation of the free layer's magnetization from the easy axis, the deviation the moment the current pulse is applied, discussed further below. 

The threshold currents for switching are (c.f.~\cite{Koch2004}):
\begin{eqnarray}
I_c^{P\rightarrow AP} &=& \frac{4 e}{\hbar P} \mu_0 M_s V \alpha \left(H_k + M_s/2 \right) \\
\label{eq:3}
I_c^{AP\rightarrow P} &=& \frac{4 e}{\hbar P} \mu_0 M_s V \alpha \left(H_k + M_s/2 \right)/ \Lambda^2,
\label{eq:4}
\end{eqnarray}
where $\alpha$ is the damping and $H_k$ is the easy axis anisotropy field.  
Important for our analysis, Eqs.~\ref{eq:1} and~\ref{eq:2} are each are of the form
\begin{equation}
I - I_c = \frac{I_c \tau_0}{\tau},
\label{eq:5}
\end{equation}
where $\tau_0 = \ln(\pi/(2\theta_0))/(\gamma \alpha (H_k + M_s/2))$
is independent of the switching direction. We therefore fit the experimental data in Fig.~\ref{Fig:Fig3} with Eq.~\ref{eq:5} under the constraint that $\tau_0$ is the same for both P$\rightarrow$AP and  AP$\rightarrow$P switching directions. The fits are displayed as cyan lines in Fig.~\ref{Fig:Fig3} and the corresponding fit parameters are listed in Table 1.

From this analysis we draw the following conclusions. First, taking the ratios of fit parameters $I_c^{P\rightarrow AP}$ to $I_c^{AP\rightarrow P}$ we find that the spin-transfer torque asymmetry is significantly reduced by diluting the free layer with Cu: for the Py case $\Lambda$ = 1.44, while in the PyCu free layer device $\Lambda$ = 1.16. Next, we consider the effect of the dilution on the P$\rightarrow$AP switching currents and determine what this implies for the device's material parameters. To this end we note that the uniaxial in-plane anisotropy in our samples can be assumed to be entirely due to the device shape, which is designed to be the same for each device (up to fabrication-induced sample to sample variations, of course). 
Comparing the P$\rightarrow$AP switching currents between the two devices, we obtain the  relation
\begin{equation}
\frac{I_c^{P\rightarrow AP,u}}{I_c^{P\rightarrow AP,d}} = \chi^2 \frac{P_d \alpha_u}{P_u \alpha_d},
\label{eq:7}
\end{equation}
where $\chi$ denotes the ratio of the saturation magnetizations $M_\mathrm{s}^u/M_\mathrm{s}^d$. The labels $u$ and $d$ stand for undiluted and diluted, respectively. Vibrating sample magnetometry (VSM) measurements give $\chi = 3.6$ (see Table 1) and thus from Eq.~\ref{eq:7} we find $\frac{P_u \alpha_d}{P_d \alpha_u}=7.6$. 

Further analysis gives estimates of the ratio of the spin polarizations in the different free layer devices and also an estimate of the ratio of the damping parameters. This can be achieved by observing that
\begin{equation}
\frac{I_c^{P\rightarrow AP,u}}{I_c^{P\rightarrow AP,d}} \frac{\tau_0^u}{\tau_0^d} \approx \frac{P_d \ln\left(\frac{\pi}{2 \theta_0^u}\right)}{P_u \ln\left(\frac{\pi}{2 \theta_0^d}\right)}\chi,
\label{eq:8}
\end{equation}
where the $\approx$, refers to the assumption that the gyromagnetic ratios do not vary between the devices. The left-hand side of Eq.~\ref{eq:8} is obtained from the fits to the experimental data and is approximately equal to 4.21. The two initial angles $\theta_0^{u(d)}$ are expectation values that depend on the device shape, the respective saturation magnetizations, and the temperature. With the saturation magnetizations in Table 1, the right-hand side of Eq.~\ref{eq:8} can be used to estimate the ratio of the spin polarizations $P_d/P_u$.  To make this estimate, we assume a Boltzmann distribution of the initial magnetization state of the free layer to obtain
\begin{equation}
\langle\theta_0^{u(d)}\rangle \approx \sqrt{\frac{\pi DkT}{2\mu_0\left(M_s^{u(d)}\right)^2V}},
\label{eq:9}
\end{equation}
where $k$ is the Boltzmann constant and $D = \frac{M_\mathrm{s}^{u(d)}}{H_\mathrm{k}^{u(d)}}\approx$ 19.6 only depends on the device shape and is assumed to be sufficiently similar for the two samples. The same applies to the device volume $V$. Inserting Eq.~\ref{eq:9} into Eq.~\ref{eq:8}, we obtain $P_d/P_u=0.85$ for $T = 3.2$ K. The value depends only weakly on the assumed temperature, ranging from 0.85 at 3.2\,K to about 0.82 at 10\,K. As a consistency check, we calculate $\langle \theta_0^{u(d)} \rangle$ = 0.015 (0.054) which are small enough for Eq.~\ref{eq:9} to be a good approximation. The above range of values is consistent with the  reduced magnetoresistance observed in the dilute free layer devices (c.f. Fig.~\ref{Fig:Fig2}). Finally, we can revisit Eq.~\ref{eq:7} to estimate $\frac{\alpha_d}{\alpha_u} \approx 6.5$, indicating about a six-fold increase of the damping due to the dilution. This is a large increase, but not entirely unexpected. Mathias {\it et al.}~\cite{Mathias2012} found a factor of three increase in the damping with 40\% Cu dilution of Py at room temperature.  Also, Rantschler {\it et al.}~\cite{Rantschler2007} found that the damping of Py at room temperature increases by  0.2$\times$10\textsuperscript{-3} per atomic percent of Cu. The analysis and particularly the very large apparent increase in damping may also be associated with the lower magnetization and exchange stiffness in the PyCu opening other dissipation channels in spin-torque switching, such as the excitation of spin-waves, or the formation micromagnetic structure in the switching process.

In summary, we have studied nanosecond switching phase diagrams for spin-valve nanopillars with in-plane magnetized PyCu and Py free layers at low temperature. The PyCu free layer sample exhibits reduced switching currents for the parallel to antiparallel configuration and significant speed-up of the characteristic switching time compared to the Py free layer device. This results in greatly reduced switching energies ($E = RI^2\tau$) for the PyCu free layer device. While the switching energy for the antiparallel to parallel configuration is reduced from 53 to 18\,fJ, the switching energy for the opposite switching direction shows over a seven-fold decrease for the diluted sample from 230\,fJ to 32\,fJ.

The clear reduction in the energy consumption of the PyCu free layer device as well as its speed-up in the switching characteristics makes it especially interesting as a low-energy data storage solution for superconducting computing. Further, our modeling suggest a means to further significant reductions in the switching energy and increases in device performance metrics. Foremost, larger reductions in switching energy require low magnetization density materials with larger spin polarization and lower damping (for example, Heusler alloys~\cite{Kubota2009,Carey2011,Andrieu2016}), which would have the added benefit of increasing the device magnetoresistance while reducing the switching current.   

\begin{acknowledgments}
We thank Jamileh Beik Mohammadi for comments on the manuscript.
We thank Canon ANELVA for providing the layer stacks and Spin Memory for patterning the devices. The research is based on work supported by the Office of the Director of National Intelligence (ODNI), Intelligence Advanced Research Projects Activity (IARPA), via contract W911NF-14-C0089. The views and conclusions contained herein are those of the authors and should not be interpreted as necessarily representing the official policies or endorsements, either expressed or implied, of the ODNI, IARPA, or the U.S. Government. The U.S. Government is authorized to reproduce and distribute reprints for Governmental purposes notwithstanding any copyright annotation thereon. This document does not contain technology or technical data controlled under either the U.S. International Traffic in Arms Regulations or the U.S. Export Administration Regulations. 
\end{acknowledgments}

%

\end{document}